\documentclass[pmlr]{jmlr}

\usepackage{layouts}

\usepackage{rotating}
\usepackage{nicematrix}
\usepackage{longtable}
\usepackage{setspace}
\usepackage{footnote}
\usepackage{threeparttable}
\usepackage{booktabs}
\usepackage[load-configurations=version-1]{siunitx} 
\usepackage{cite}
\usepackage{blindtext}
\usepackage{hyperref}
\usepackage{tablefootnote}
\usepackage{longtable}

\newcolumntype{C}[1]{>{\centering\arraybackslash}p{#1}}

\DeclareRobustCommand{\sbseries}{\fontseries{sb}\selectfont}
\DeclareTextFontCommand{\textsb}{\sbseries}

\theorembodyfont{\upshape}
\theoremheaderfont{\scshape}
\theorempostheader{:}
\theoremsep{\newline}

\jmlrvolume{1}
\jmlryear{2022}
\jmlrworkshop{HEAR 2021}

\title[BYOL-S: Learning Self-supervised Speech Representations by Bootstrapping]{BYOL-S: Learning Self-supervised Speech Representations by Bootstrapping}

  \author{\Name{Gasser Elbanna\nametag{\thanks{GE and KEH performed this work as interns at Logitech.}}} \Email{gasser.elbanna@epfl.ch}  \\ 
  \addr Ecole Polytechnique Fédérale de Lausanne, Lausanne, Switzerland \\
   \Name{Neil Scheidwasser-Clow} \Email{neil.scheidwasser-clow@epfl.ch} \\ 
   \addr Ecole Polytechnique Fédérale de Lausanne, Lausanne, Switzerland\\
   \Name{Mikolaj Kegler}\Email{mikolaj.kegler16@imperial.ac.uk} \\ 
   \addr Imperial College London, London, United Kingdom \\
   \Name{Pierre Beckmann}\Email{pierre.beckmann@unil.ch} \\
   \addr Université de Lausanne, Lausanne, Switzerland \\
   \Name{Karl El Hajal\nametag{\footnotemark[1]}}\Email{karl.elhajal@epfl.ch} \\
   \addr Ecole Polytechnique Fédérale de Lausanne, Lausanne, Switzerland \\
   \Name{Milos Cernak}\Email{milos.cernak@ieee.org} \\
   \addr Logitech Europe S.A., Lausanne, Switzerland \\
   }

\begin{document}

\maketitle

\begin{abstract}
    Methods for extracting audio and speech features have been studied since pioneering work on spectrum analysis decades ago. Recent efforts are guided by the ambition to develop general-purpose audio representations. For example, deep neural networks can extract optimal embeddings if they are trained on large audio datasets. This work extends existing methods based on self-supervised learning by bootstrapping, proposes various encoder architectures, and explores the effects of using different pre-training datasets. Lastly, we present a novel training framework to come up with a \textit{hybrid} audio representation, which combines handcrafted and data-driven learned audio features. All the proposed representations were evaluated within the HEAR NeurIPS 2021 challenge for auditory scene classification and timestamp detection tasks. Our results indicate that the hybrid model with a convolutional transformer as the encoder yields superior performance in most HEAR challenge tasks.
\end{abstract}

\begin{keywords}
audio embeddings, representation learning, self-supervised learning, hybrid representations
\end{keywords}


\section{Introduction}

Humans are able to learn, memorize, and distinguish various auditory patterns from limited data by projecting low-level audio inputs to high-level representations in the brain~\citep{griffiths1999human}. Inspired by human capabilities, a substantial body of research has been dedicated over the past few decades to build models capable of extracting and representing auditory information. Historically, handcrafted feature sets, based on digital signal processing (DSP), have been employed to extract audio representations~\citep{liu1998audio, eyben2010opensmile}. However, the recent success of deep learning in computer vision and natural language processing has propelled the development of data-driven frameworks, where deep neural networks (DNNs) are trained on large audio corpora to capture crucial features~\citep{hershey2017cnn,van2018representation}.

Two main methods currently coexist to build deep-learning-based audio representations: supervised~\citep{hershey2017cnn, beckmann2021word} and
self-supervised learning~\citep{baevski2020,niizumi2021byol}. While supervised methods have been at the center of most initial breakthroughs in vision, audio, and language understanding, they are inevitably limited by their reliance on well-defined labels for each training data input. Conversely, self-supervised learning aims at leveraging relations within input data to generate pseudo-labels, thus creating proxy supervised tasks \citep{liu2021self, murphy2022}. Recently, several self-supervised models have been proposed as robust general-purpose audio representations \citep{van2018representation, shor2020towards, saeed2021contrastive, niizumi2021byol, shor2021universal}. Most models are trained using contrastive learning setups, where an encoder network learns to produce a latent space representation by assessing the degree of similarity between input examples \citep{van2018representation, shor2020towards, saeed2021contrastive}. In this framework, similar inputs should be mapped closer in the latent space, whereas unrelated examples should appear more distant. In the context of audio, \textit{similarity} can be measured in terms of temporal proximity \citep{shor2020towards} or, more simply, whether two audio segments originate from the same source or not \citep{saeed2021contrastive}.

However, \citet{niizumi2021byol} argue that contrastive learning frameworks may potentially suffer from several limitations when it comes to audio representation learning. For example, similar rhythmic patterns can be found in different audio sources. Alternatively, short impulsive sounds, such as glass breaking, may be found only in a single example of an audio clip and thus appear as “dissimilar” to other inputs from the same clip. Consequently, \citet{niizumi2021byol} proposed Bootstrap Your Own Latent for Audio (BYOL-A) to learn audio representations by comparing augmented views of a single audio segment. Inspired by the success of BYOL for self-supervised image representation \citep{grill2020}, BYOL-A achieved competitive results in various tasks, including speaker identification, language identification, speech commands, and musical instrument classification.

That being said, handcrafted DSP-based feature sets remain widely used in various speech- and music-related applications. For instance, extensive feature sets such as openSMILE~\citep{eyben2010opensmile} often constitute a strong baseline in paralinguistic tasks and challenges \citep{schuller2013interspeech, schuller2016interspeech, schuller2020interspeech}. Unlike DNNs, such frameworks are completely transparent and interpretable. Moreover, they do not require any training, thus reducing both computational costs and risks of model overfitting (omitting any bias introduced by the handcrafted feature set designer). While the above-outlined properties make the DSP-based features remain relevant for many applications, most recent pre-trained DNNs significantly outperform handcrafted feature sets \citep{shor2020towards, scheidwasser2021serab}.
Accordingly, in this paper, we propose new extensions of BYOL-A for speech representation learning. Whereas the original BYOL-A model was trained on the entire AudioSet \citep{gemmeke2017audio}, a large audio dataset with more than 5800 hours of audio, here we retrained different models on a speech-specific subset from AudioSet creating BYOL for speech (BYOL-S). Originally, the BYOL-S model was developed for speech emotion recognition (SER) and outperformed BYOL-A and other pre-trained models in the context of a speech emotion recognition adaptation benchmark (SERAB) \citep{scheidwasser2021serab}. Here, we extend our previous work to assess BYOL-S, as a general-purpose audio representation and to thoroughly study the impact of different hyperparameters and training protocols on the model performance. In particular, we introduce different encoder architectures (Section~\ref{sec:encoder}) than the default one used in \citet{niizumi2021byol}. With the aim of leveraging the best of both DSP- and DNN-based approaches, we finally assessed the impact of incorporating DSP-based features to the BYOL training paradigm. This led to the development of a novel pre-training protocol for BYOL-S that combines learned and fixed DSP-based handcrafted features (Section~\ref{sec:hybrid}). Such a \textit{hybrid} approach can facilitate the pre-training of the model by grounding the complex DNN features with the considerably simpler DSP-based ones.

All models were evaluated in the context of the Holistic Evaluation of Audio Representations (HEAR) 2021 challenge\footnote{\url{https://neuralaudio.ai/}}, a challenge aimed at designing general-purpose audio embeddings. Launched at NeurIPS 2021, the HEAR challenge featured a new 16-task benchmark suite to compare models, including scene-based (i.e., audio classification) and timestamp-based tasks (i.e., sound event detection). Importantly, the benchmark comprises data from a variety of sources, e.g., human speech, environment sounds, and music.

\section{Methods}
\label{sec:methods}
\begin{figure}[htbp]
    \includegraphics[width=0.9\textwidth]{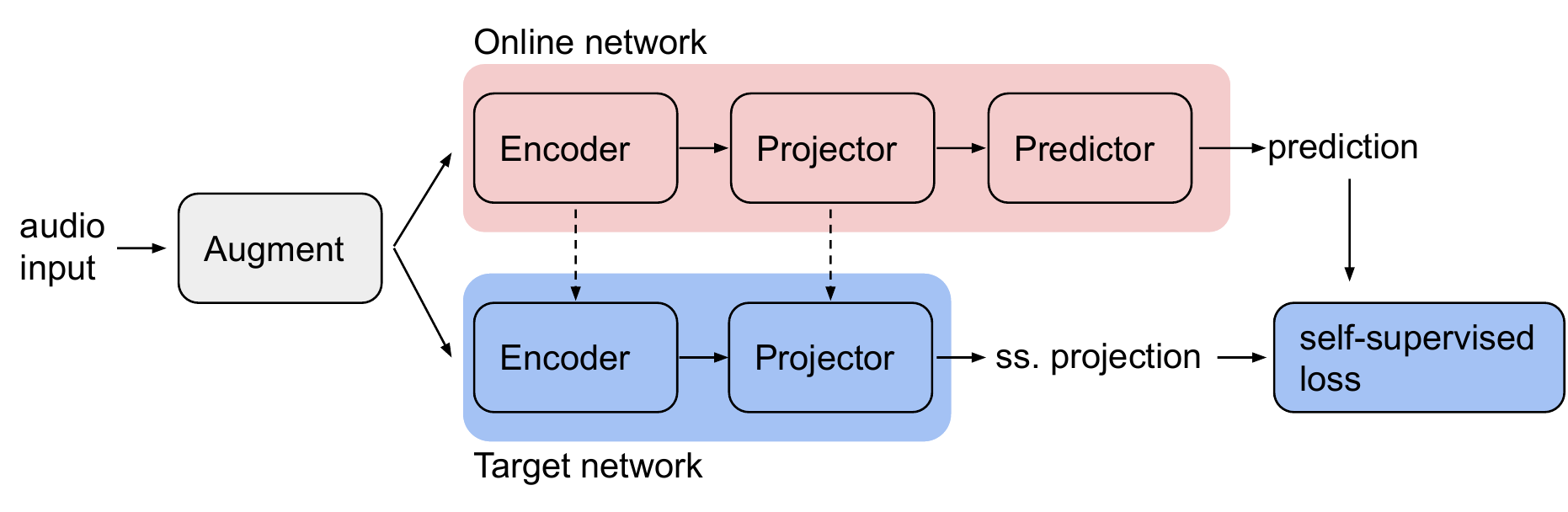}
    \caption{Schematic of the BYOL-A architecture, showing the main modules involved in the training paradigm. Adapted from \citep{grill2020, niizumi2021byol}. ss: self-supervised. The dotted lines indicate the fact that the target network parameters are updated as a moving average of the online parameters.}
    \label{fig:byola_plot}
\end{figure}

All proposed models constitute extensions of BYOL-A~\citep{niizumi2021byol}, an adaptation of \textit{bootstrap your own latent} (BYOL) \citep{grill2020} for general-purpose audio representation learning. More specifically, we extended our previous work \citep{scheidwasser2021serab} for speech representation by varying the encoder networks within the BYOL framework, as shown in Figure \ref{fig:byola_plot}. In addition, we explored \textit{hybrid} approaches by combining BYOL-like networks with hand-crafted features from openSMILE \citep{eyben2010opensmile}, with the aim of assisting the self-supervised network with spectral and prosodic information during the training process.

While contrastive learning setups typically rely on comparing different audio segments to learn representations, BYOL-A models learn by comparing two augmented versions of a single audio input~\citep{niizumi2021byol}. During pre-training, the input audio is first preprocessed into a 64-band log-mel magnitude spectrogram (LMS) to produce a 2D input. All spectrograms were generated within a frequency range of 60 to 7800 Hz, a sampling rate of 16 kHz, a window length of 25 ms, and a hop size of 10 ms. By default, the BYOL-A framework uses random 96-frame-long segments from the input LMS for model training \citep{niizumi2021byol}, approximately corresponding to 0.95 s of audio. We experimented with changing the training window size, as discussed in Section~\ref{sec:res_windows}. However, the pre-trained model can be fed with audio samples of any duration during the inference stage. The extracted log-mel spectrograms are then of dimension $64 \times T$, where $T$, the number of spectrogram frames, depends on the sample length.

Subsequently, the LMS input is fed to an augmentation module, which first standardizes the spectrogram before applying two different data augmentations: mixup \citep{zhang2018mixup}, i.e., adding randomly mixed audio samples in the background of the input, and random resize cropping, which in the context of audio spectrograms can be equated with pitch-shifting and time-stretching. The random augmentation is applied twice to produce two randomly augmented views of the input spectrogram. These views thus share the same input source signal but are processed using two slightly different audio augmentations. At last, both views are re-standardized to account for any statistical drifts.

The models are trained to predict the representation of the first augmented version from the representation of the second. To that end, the two augmented views are fed respectively to an \textit{online} and a \textit{target} network (Figure \ref{fig:byola_plot}), which have a different set of weights. The rationale behind encompassing two different networks is to build a self-supervised learning paradigm that makes the target network act as a pseudo label generator that is, then, compared to the online network output. Both networks comprise an \textit{encoder} and a \textit{projection} head: the encoder extracts a representation of the augmented input, whereas the \textit{projection} head helps to map the representation to a lower dimensional latent space with embedding size of 256. Additionally, the online network includes a predictor head to avoid collapsing solutions with constant representations \citep{grill2020}. Both the projector and predictor are two-layer feed-forward networks. Finally, a simple mean-squared error (MSE) is used as a loss function to minimize the difference between the online predictor and the target projector outputs (ss. projection). While the online parameters were updated using Adam optimization \citep{kingma2015} with a learning rate of 0.0003, target parameters were updated as an exponential moving average of the online parameters, which empirically showed to improve training stability and yield more robust embeddings \citep{grill2020}.

The default encoder used in BYOL-A is a simple convolution neural network (CNN) adapted from a solution to the Automated Audio Captioning task of the Detection and Classification of Acoustic Scenes and Events (DCASE) 2020 Challenge \citep{niizumi2021byol}. More specifically, the network comprises three convolution blocks (1 block = Conv2d-BatchNorm-ReLU-MaxPool2d) with 64 3x3 filters each, followed by two linear layers projecting the output of the final block onto a 2048-dimensional embedding.

In addition to the default encoder, we explore different encoder architectures in Section \ref{sec:encoder} and evaluate their performance across the HEAR challenge tasks.

\subsection{Encoder Architectures}
\label{sec:encoder}

Three different encoding networks were benchmarked against the HEAR tasks to gain further insight into the robustness of certain audio representations. These networks, presented in Table \ref{tab:encoders}, are: ResNetish-34 \citep{hershey2017cnn}, a convolutional LSTM inspired by \citep{passricha2020hybrid} and the convolutional vision transformer (CvT) \citep{wu2021cvt}. While both ResNetish and CLSTM networks rely on CNNs for feature extraction, CLSTMs also incorporate recurrent neural networks (RNNs) to leverage the temporal properties of audio data. Inspired by the success of transformers in computer vision, \citet{wu2021cvt} showed with CvT that including convolutional token embeddings and projections within transformer layers yielded competitive results for image classification.

\begin{table}[h!]
  \centering
  \caption{Encoder architectures studied within the BYOL-S framework.}\vspace{4pt}
  \begin{tabular}{lrr}
    \toprule
   \textsb{Encoder} & \textsb{Parameters (M)}  & \textsb{Embedding} \\
    & \textsb{} & \textsb{size} \\
    \midrule
    Default (CNN)~\citep{niizumi2021byol}             & 5.3    & 2048 \\
    ResNetish-34~\citep{hershey2017cnn} & 21.3   & 2048 \\
    CLSTM~ \citep{passricha2020hybrid}       & 18.6   & 1024 \\
    CvT~\citep{wu2021cvt}         & 5.0      & 2048 \\
    \bottomrule
  \end{tabular}
  \label{tab:encoders}
\end{table}


\subsubsection{ResNetish Model} 
An audio version of a 34-layer residual network~\citep{he2016deep} was implemented using the same modifications as~\citet{hershey2017cnn} for large-scale audio classification. In accordance with~\citet{shor2020towards}, we refer to this encoder as ResNetish-34 (Table \ref{tab:encoders}). Although ResNetish-50 was the most robust for AudioSet classification compared to other common CNNs \citep{hershey2017cnn}, we implemented a lighter version, ResNetish-34, to avoid overfitting the pre-training datasets. The implementation was derived from an adaptation of ResNetish in \texttt{PyTorch}\footnote{\url{https://github.com/daisukelab/sound-clf-pytorch}} with a final embedding size of 2048.

\subsubsection{Convolution with LSTM} 


The ability of RNNs such as LSTM-based networks to capture temporal dependencies in sequential data led us to implement a bidirectional LSTM (BiLSTM)~\citep{graves2005} network with CNN features (CLSTM) to explore the benefits of RNNs. This model is inspired by previous work by ~\citet{passricha2020hybrid} on automatic speech recognition (ASR). Their model comprised two 256-filter convolution layers (with 9x9 and 4x3 (frequency x time) kernels), followed by two BiLSTM layers and three feedforward layers to generate higher-order representations. While the CNN architecture was similar to the original implementation, we only used one 512-layer BiLSTM layer and one 1024-dimensional fully connected layer to prevent overfitting during pre-training. 



\subsubsection{Convolution Vision Transformer} 
Inspired by transformer architectures for vision \citep{dosovitskiy2020}, \citet{wu2021cvt} proposed the Convolutional vision Transformer (CvT), which leverages the advantages of both CNNs (i.e., detecting fine-grained local patterns) and Transformers (i.e., learning long-range global context). The network architecture comprises three stages. Each stage comprises a convolutional token embedding layer and a convolutional transformer block, the latter of which includes a convolutional projection layer followed by multi-head self-attention. We adapted the implementation from\footnote{\url{https://github.com/lucidrains/vit-pytorch}} and built a lightweight version of CvT with one transformer block per stage with embedding sizes of 64, 256, and 512, respectively. Using \textit{mean+max} temporal aggregation, the final layer outputs a vector of dimensions 2048. Such an encoder yielded competitive results in SER tasks \citep{scheidwasser2021serab}.

\subsection{Hybrid Representations}
\label{sec:hybrid}
Despite current progress in deep learning-based audio and speech representation, handcrafted (DSP-based) feature sets remain competitive in various paralinguistic challenges \citep{schuller2013interspeech, schuller2020interspeech}. This motivated us to study the benefits of combining DSP-based features and data-driven features by adding a third \textit{supervision} module to the BYOL-S framework \citep{elbanna22_interspeech}. Here, the module simply consists of features extracted using the ComParE 2016 acoustic feature set \citep{schuller2013interspeech} from openSMILE (OS) \citep{eyben2010opensmile}\footnote{\url{https://audeering.github.io/opensmile-python}}. This extensive feature set comprises 6373 static features, including acoustic functionals, low-level descriptors (LLDs) and LLD derivatives.

\begin{figure}[ht]
        \centering
        \includegraphics[width=\textwidth]{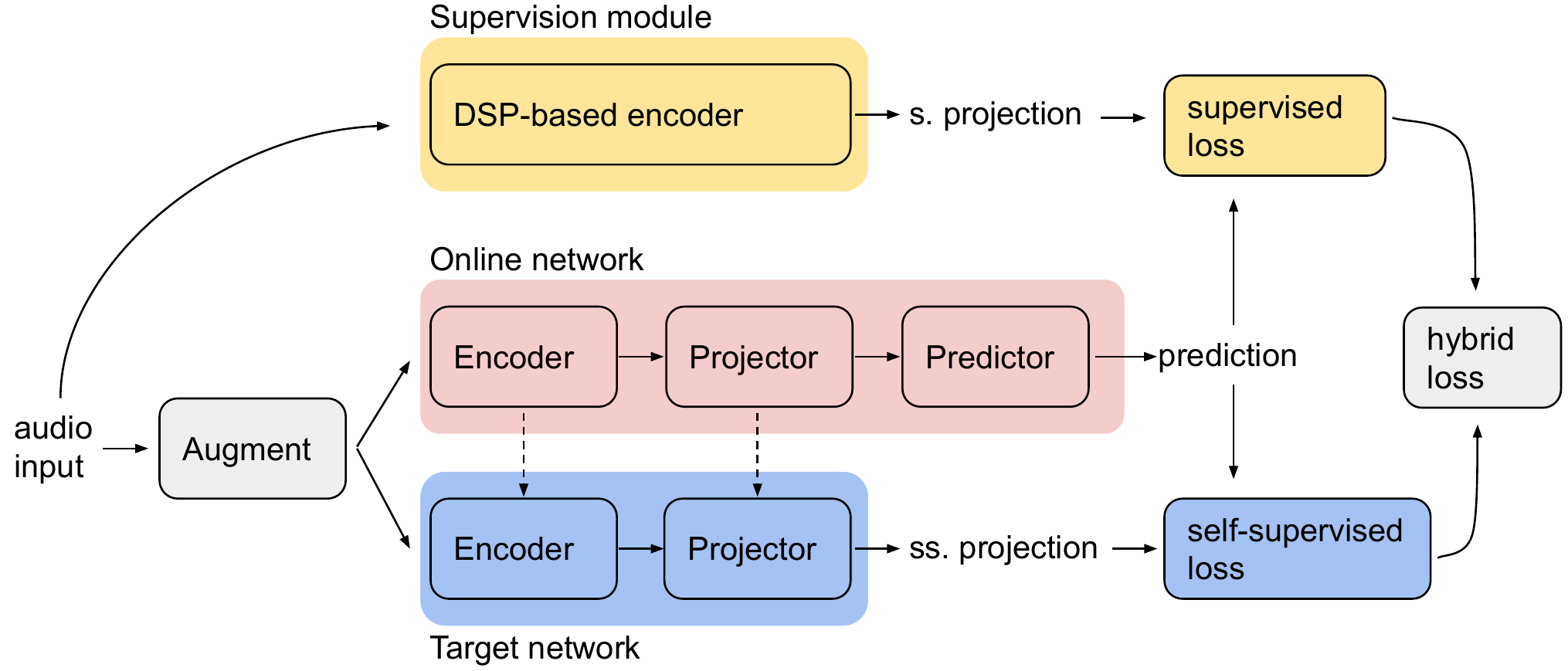}
        \caption{The hybrid BYOL-S framework leverages both self-supervised features and handcrafted features from openSMILE. $s$: supervised, $ss$: self-supervised. The dotted arrows indicate that the target parameters are updated as a moving average of the online parameters.}
        \label{fig:hybrid_byols_plot}\vspace{-1em}
\end{figure}

We denote the resulting model as \textit{Hybrid BYOL-S}, since the online network learns to strike a balance between self-supervised, learned features and supervised features from a fixed feature set, as illustrated in Figure \ref{fig:hybrid_byols_plot}. More specifically, the model is trained to optimize a sum of two loss functions: the \textit{self-supervised} BYOL loss between the online and target outputs ($\mathcal{L}_{ss}$), and a second \textit{supervised} loss ($\mathcal{L}_{sup}$), computed as the mean squared error between the outputs of the online and DSP-based networks. To get a more exhaustive view of the performance of the hybrid model, we finally explored different weights $\alpha$ and $\beta$ for $\mathcal{L}_{sup}$ and $\mathcal{L}_{ss}$, respectively, leading to the following hybrid loss:

\vspace{-0.25cm}
\begin{equation}
	\mathcal{L}_{hybrid} = \alpha \mathcal{L}_{sup} + \beta 
	\mathcal{L}_{ss}
	\vspace{-0.1cm}
\label{eq:loss}
\end{equation}
Note that openSMILE features were standardized before computing the supervised loss and the output dimension of the online and target networks' projectors was changed from 256 to 6373 to accommodate the size of the openSMILE features. However, the embedding size of the encoders (2048) remained unchanged.

\vspace{-0.15cm}

\subsection{Implementation Details}
For all experiments, we used the same pre-training hyperparameters used in~\citet{niizumi2021byol}\footnote{\url{{https://github.com/nttcslab/byol-a}}}, in which models were trained for 100 epochs using Adam and a batch size of 256 with a learning rate of 0.0003. Pre-trained models and code for embedding computation (explained in Section~\ref{sec:downstream}) are available at\footnote{\url{{https://github.com/GasserElbanna/serab-byols}}}. 

\subsection{Embedding Generation and Downstream Evaluation}
\label{sec:downstream}
All 16 tasks from the HEAR challenge were used to evaluate the robustness of our pre-trained models, which include both scene-based and timestamp-based tasks. Scene-based tasks (Table \ref{tab:scene_datasets}) consist of multi-class or multi-label audio classification and can be further divided into three categories: speech, environmental sounds, and music. Although containing human and musical sounds, FSD50K was included in the “environmental sounds” category. On the other hand, timestamp-based tasks (Table \ref{tab:timestamp_data}) consist of sound event detection or transcription and include two datasets: MAESTRO and Task 2 from DCASE 2016.

All audio clips were sampled at 48 kHz\footnote{\url{https://doi.org/10.5281/zenodo.5887964}}. For each dataset, all recordings were resampled to 16 kHz using \texttt{torchaudio}\footnote{\url{https://pytorch.org/audio/stable/transforms.html\#torchaudio.transforms.Resample}} to align with our model implementation. Subsequently, embeddings were generated from pre-trained models differently depending on the task type. For scene-based tasks, the entire audio samples were used to produce the embedding, while timestamp-based tasks required the sample to be chunked into fixed-size segments before generating window representations aligned with their corresponding timestamps. 
Finally, the embeddings were evaluated using the \texttt{hear-eval}\footnote{\url{https://github.com/neuralaudio/hear-eval-kit}} toolkit. This toolkit trains a shallow fully-connected predictor on a train set of a downstream task, which is optimized on a validation set before evaluation on an unseen test set. Despite slight discrepancies between the official challenge results and the results presented herein due to different resampling methods, the produced scores remained stable and consistent with \citet{turian2022}. We observe however a higher discrepancy for timestamp-related tasks, especially MAESTRO, which we ascribe to higher sensitivity to resampling methods.  

\clearpage
\setlength{\tabcolsep}{2pt}
\begin{table}[h!]
\small
\centering
\caption{HEAR scene-based tasks and datasets. clf.: classification. Adapted from \citet{turian2022}.}
\begin{tabular}{lp{3.9cm}rr}
\toprule
 \textsb{Dataset}                       & \textsb{Task}                            & \textsb{\# Classes}                            & \textsb{Total}          \\
 & & & \textsb{duration (h)} \\
\midrule
\multicolumn{4}{c}{Speech}\\
\midrule
CREMA-D \citep{cao2014crema}                 & Emotion recognition                            & 6                            & 5.3                          \\
LibriCount \citep{stoter2018libricount}              & Speaker count estimation           & 11                           & 7.9                          \\
Speech Commands 5h/full  \citep{warden2018speech}& Keyword spotting                   & 12                           & 6.4/27.9 \\
Vocal Imitations \citep{kim2018vocal}        & Vocal imitation clf.           & 302                          & 17.5                         \\
VoxLingua \citep{valk2021voxlingua107}               & Language identification            & 10                           & 5.0                            \\
\midrule
\multicolumn{4}{c}{Environment sounds}\\
\midrule
Beehive states \citep{nolasco2019audio}          &  Beehive identification                                  & 2& 96                           \\
ESC-50 \citep{piczak2015dataset}                  & Sound classification & 50                          & 2.8                          \\
FSD50K \citep{fonseca2020fsd50k}                  & Sound classification         & 200                        & 108.3                        \\
Gunshot triangulation \citep{cooper_gunshots_2020}  &  Gunfire location                                  & 4                            & 0.04                         \\
\midrule
\multicolumn{4}{c}{Music}\\
\midrule
Beijing Opera \citep{mi_tian_2014_1285212}           & Instrument classification         & 4                            & 0.31                         \\
GTZAN (Genre)   \citep{tzanetakis2002musical}    & Genre classification         & 10                           & 8.3                          \\
GTZAN (Speech/Music)\tablefootnote{\url{http://marsyas.info/downloads/datasets.html\#music-speech}}& Speech vs music clf.    & 2                             & 1.07                         \\
Mridangam \citep{anantapadmanabhan2013modal}                   & Stroke \& tonic clf. & 10 \& 6 & 1.6                          \\
NSynth 5h/50h \citep{engel2017neural}           & Pitch \& chroma clf.       & 88 \& 12 & 5.6/54.5 \\
\bottomrule
\end{tabular}
\vspace{-0.5cm}
\label{tab:scene_datasets}
\end{table}
\setlength{\tabcolsep}{6pt}

\begin{table}[h!]
\caption{HEAR timestamp-based tasks and datasets.}
\small
\centering
\begin{tabular}{llr}
\toprule
\textsb{Dataset}           & \textsb{Task}  & \textsb{Total} \\
 & & \textsb{duration (h)} \\
\midrule
MAESTRO \citep{hawthorne2018enabling}           & Music transcription     & 6.2                \\
DCASE 2016 Task 2 \citep{Mesaros2018_TASLP}     & Office sound detection  & 2.4                 \\
\bottomrule
\end{tabular}
\vspace{-0.4cm}
\label{tab:timestamp_data}
\end{table}


\section{Results}
Tables \ref{tab:speech}, \ref{tab:env} and \ref{tab:music} present the results for scene-based tasks (Table \ref{tab:scene_datasets}) pertaining to speech, environmental sounds, and music, respectively. Test accuracy (in percentage) was used as the evaluation metric for each task except FSD50K (mAP). The best scores are shown in bold. For comparison, each set of experiments within each table is evaluated against the scores obtained by two HEAR baseline models:

\begin{itemize}
    \setlength\itemsep{0.5em}
    \item \texttt{wav2vec2} \citep{baevski2020}, a self-supervised framework pretrained on 100K hours of speech from VoxPopuli \citep{wang2021voxpopuli}. The model comprises a 1D convolutional feature encoder followed by a positional transformer for context representation.
    \item \texttt{CREPE} \citep{kim2018crepe}, a 1D CNN for pitch estimation pre-trained on 16 hours of synthesized music.
\end{itemize}

\subsection{Contribution of pre-training datasets: BYOL-A, BYOL-S and BYOL-S\texttt{++}}
First, we considered re-training BYOL-A from scratch with different pre-training datasets. Whereas BYOL-A was pre-trained on AudioSet~\citep{gemmeke2017audio}, our early submission (BYOL-S) was only pre-trained on speech samples of AudioSet. Moreover, we introduce BYOL-S\texttt{++}, trained on LibriSpeech \citep{panayotov2015librispeech} in addition to the speech subset of AudioSet. With approximately 960 hours of data (Table \ref{tab:training_sets}), enriching the pre-training corpus with LibriSpeech samples further diversifies the format of speech samples from which the model could learn. Whereas AudioSet mostly features spontaneous speech surrounded by environmental sounds or music, LibriSpeech consists of read English speech derived from audiobooks recorded in a studio environment. All results pertaining to these experiments are shown in the \textsb{Pre-training dataset} section of Tables \ref{tab:speech}, \ref{tab:env} and \ref{tab:music}.


\begin{table}[h!]
  \small
  \centering
  \caption{Datasets used for self-supervised training}\vspace{4pt}
  \begin{tabular}{llr}
    \toprule
    \textsb{Model Name} & \textsb{Dataset} & \textsb{Duration (h)} \\
    \midrule
     BYOL-A & AudioSet & 5800   \\
     BYOL-S & AudioSet (Speech subset) & 2190 \\
     BYOL-S\texttt{++} & AudioSet (Speech subset) + LibriSpeech & 3150 \\
     \bottomrule
  \end{tabular}
  \label{tab:training_sets}
\end{table}

\subsection{Contribution of the pre-training window size}
\label{sec:res_windows}
All models were originally pre-trained with random 96-frame segments of audio spectrograms, corresponding to 0.95 s of audio. To study the effect of this window length, we compared four versions of BYOL-S, pre-trained with a window size of 0.5, 0.95, 1.425 and 2~s, respectively. As mentioned in Section \ref{sec:methods}, all models could still be fed with samples of any duration for downstream evaluation. The \textsb{BYOL-S window size (s)} section of Tables \ref{tab:speech}, \ref{tab:env} and \ref{tab:music} shows the results obtained by training BYOL-S with the window sizes mentioned above. To reduce training time, the batch size was reduced from 256 to 128 for these experiments. 


\subsection{Comparison of encoder architectures: ResNetish-34, CLSTM, CvT}
\label{sec:res_encoders}
The \textsb{BYOL-S encoder} section of Tables \ref{tab:speech}, \ref{tab:env} and \ref{tab:music} shows the results obtained by replacing the default encoder in BYOL-S with the DNNs described in Section \ref{sec:encoder}. The other pre-training parameters were unchanged.

\subsection{Hybrid versions of BYOL-S and BYOL-S/CvT}
Here, we evaluated the performance of hybrid models (Figure \ref{fig:hybrid_byols_plot}), obtained by combining handcrafted and learnable feature sets during model pre-training. Following Section \ref{sec:res_encoders}, two models were evaluated against the HEAR benchmark: Hybrid BYOL-S (using the "default" BYOL-S encoder) and Hybrid BYOL-S/CvT (i.e., with CvT encoding).

Moreover, to assess the impact of self-supervised and supervised losses on overall performance, we tested different values for loss weights $\alpha$ and $\beta$ (Eq. \ref{eq:loss}) when pre-training the hybrid BYOL-S/CvT. All results pertaining to these experiments are shown in the \textsb{Hybrid models} and \textsb{Hybrid BYOL-S/CvT} sections of Tables \ref{tab:speech}, \ref{tab:env}, and \ref{tab:music}.

Lastly, we validated the relevance of the hybrid training protocol against embeddings produced using only openSMILE features and the concatenation of the latter with BYOL-S/CvT features. Here, we modified the embedding generation scripts in the \texttt{hear-eval} toolkit by applying per-fold standardization to account for large values produced by openSMILE (mainly due to the computation of LLD functionals and deltas), which hindered convergence during downstream training. All results pertaining to these experiments are shown in the \textsb{Hybrid} section of Tables \ref{tab:speech}, \ref{tab:env} and \ref{tab:music}.

\begin{table}[h!]
\small
\centering
\caption{Top-1 accuracy of all proposed models on speech-related tasks. + denotes concatenation.}
    \begin{tabular}{l@{\hspace{20pt}}cccccc@{\hspace{20pt}}c}
    \toprule
        & \rotatebox{90}{CREMA-D} 
        & \rotatebox{90}{LibriCount} 
        & \rotatebox{90}{\parbox{.25cm}{Speech \\ Commands (5h)}} 
        & \rotatebox{90}{\parbox{.25cm}{Speech \\ Commands (all)}} 
        & \rotatebox{90}{\parbox{.25cm}{Vocal \\ Imitations}} 
        & \rotatebox{90}{VoxLingua} 
        & \rotatebox{90}{Average} \\
    \midrule
    \textsb{HEAR baselines:} \\
    CREPE & 36.2 & 49.9 & 16.8 & 19.6 & 5.1 & 15.1 & 23.8 \\
    wav2vec2 & 65.7 & 67.6 & 79.7 & 88.5 & 7.2 & \textsb{49.7} & 59.7 \\
    \midrule
    \textsb{Pre-training dataset:}\\
    BYOL-A            & 62.3          & 78.8        & 89.6 & 92.4 & 13.7          & 39.0          & 62.6 \\
    BYOL-S            & \textsb{66.4} & 78.5        & 92.6          & 94.3 & 15.1          & 41.2          & 64.7 \\
    BYOL-S\texttt{++} & \textsb{66.4} & \textsb{80.0} & \textsb{93.2} & \textsb{95.0}  & \textsb{15.4} & 47.8 & \textsb{66.3} \\
    \midrule
    \textsb{BYOL-S window size (s):}\\
    0.5   & \textsb{66.5} & \textsb{86.0}          & 91.2             & 93.4          & 14.2         & 43.0           & 64.8    \\
    0.95  & 65.5          & 83.0          & 91.9             & \textsb{94.5} & 14.8         & 44.4          & 65.2     \\
    1.425 & 65.7          & 81.0          & 90.0               & 93.1          & \textsb{16.0} & 43.1          & 64.7    \\
    2     & 65.6          & 81.8 & \textsb{92.2}    & 93.4          & 14.3         & 47.3 & \textsb{65.8}    \\
    \midrule
    \textsb{BYOL-S encoder:}\\
    Default           &     \textsb{66.4}     & 78.6 & \textsb{92.6}    & \textsb{94.3}              & 15.1           & 41.2 & 64.7          \\
    Resnetish-34 & 63.5 & 77.0           & 83.2    & 88.9              & 13.3           & 35.8 & 60.3         \\
    CLSTM       & 64.0 & 78.1 & 91.3   & 92.5   & 11.9  & 24.3 & 60.4  \\
    CvT          & \textsb{66.4}          & \textsb{84.8}          & 92.0     & 93.1              & \textsb{16.2}           & 37.0 & \textsb{64.9}\\
    \midrule
    \textsb{Hybrid models:}\\
    openSMILE (OS) only & 59.4 & 66.7 & 70.1 & 80.2 & 13.1 & 25.1 & 52.4\\
    BYOL-S/CvT + OS & 65.4 & 82.7 & 77.6 & 86.9 & \textsb{17.4} & 29.5 & 59.9 \\
    Hybrid BYOL-S/Default   & 66.1 & 81.6 & 91.6 & 93.8 & 14.2 & 43.8 & 65.2 \\
    Hybrid BYOL-S/CvT       & \textsb{67.2} & \textsb{83.5} & \textsb{92.6} & \textsb{95.8} & 16.3 & 42.2 & \textsb{66.3} \\
    \midrule
    \textsb{Hybrid BYOL-S/CvT:}\\
    \textsb{$\alpha$:$\beta$ ratio}\\
    1:4 & 62.9          & \textsb{84.4} & 86.2          & 90.7          & 12.0          & 27.1 & 60.6\\
    1:2 & 64.4          & 83.5          & 89.6          & 92.7          & 13.6          & 35.3 & 63.2\\
    2:3 & 66.0          & 84.1 & 91.9          & 93.8          & 14.6          & 38.7 & 64.9\\
    1:1 & \textsb{67.2} & 83.5          & 92.6 & \textsb{95.8} & \textsb{16.3} & 42.2 & \textsb{66.3}\\
    3:2 & 66.7 & 82.7          & 92.5          & 94.8          & 15.2          & 34.0 & 64.3\\
    2:1 & 65.8          & 81.2          & 92.7 & 94.7          & 15.8 & 39.8 & 65.0\\
    4:1 & 66.5          & 80.6          & \textsb{93.1} & 94.6          & 16.2 & 38.0 & 64.8\\
    \bottomrule
    \end{tabular}
    \label{tab:speech}\vspace{-0.25cm}
\end{table}

\begin{table}[htbp]
    \small
    \centering
    \caption{Performance on environmental sound-related datasets. Top-1 accuracy was used for each dataset but FSD50K (mAP). Due to exhausting GPU memory problems for several models (shown with superscript $^*$), the reported average does not include \textit{Beehive states}.}
    \begin{tabular}{l@{\hspace{20pt}}cccc@{\hspace{20pt}}c}
    \toprule
        & \rotatebox{90}{Beehive states} 
        & \rotatebox{90}{ESC} 
        & \rotatebox{90}{FSD50K} & \rotatebox{90}{\parbox{.25cm}{Gunshot \\ triangulation}}       
        & \rotatebox{90}{Average}\\
    \midrule
    \textsb{HEAR baselines:} \\
    CREPE & 50.4 & 29.4 & 15.9 & 91.7 & 45.7\\
    wav2vec2 & -$^*$ & 59.2 & 34.6 & 77.1 & 57.0\\
    \midrule
    \textsb{Pre-training dataset:}\\
    BYOL-A            & 48.8          & 78.9          & 48.9          & 87.5          & 71.8          \\
    
    BYOL-S            & 52.8 & \textsb{81.9} & \textsb{49.9} & \textsb{96.4} & \textsb{76.1} \\
    
    BYOL-S\texttt{++} & \textsb{55.0}          & 80.0            & 49.4          & 92.9          & 74.1 \\
    
    \midrule
    \textsb{BYOL-S window size (s):}\\
    0.5   & \textsb{59.7} & \textsb{83.3} & 49.5         & 89.3          & 74.0\\
    0.95  & 57.5          & 81.4          &  \textsb{50.0} & \textsb{95.2} & \textsb{75.5}\\
    1.425 & 56.5          & 80.3          & 49.9         & 89.3          & 73.2\\
    2     & 53.5          & 81.3          & 49.6         & \textsb{95.2} & 75.4 \\
    \midrule
    \textsb{BYOL-S encoder:}\\
    Default &  \textsb{52.8} & \textsb{81.9} & \textsb{49.9} & \textsb{96.4} & \textsb{76.1 }\\
    Resnetish-34 & \textsb{52.8}  & 71.5 & 43.3 & 86.3 & 67.0 \\
    CLSTM &  51.4 & 73.9 & 39.4 & 86.9 & 66.7 \\
    CvT & -$^*$ & 79.9 & 48.3 & 89.3 & 72.5\\
    \midrule
    \textsb{Hybrid models:}\\
    openSMILE (OS) only & -$^*$ & 68.2 & 34.5 & \textsb{98.8} & 67.2\\  
    BYOL-S/CvT + OS & -$^*$ & 76.9 & 44.2 & \textsb{98.8} & 73.3\\
    Hybrid BYOL-S/Default & \textsb{53.0}  & 82.4 & 48.9 & 95.2 & 75.5 \\
    Hybrid BYOL-S/CvT     & -$^*$  & \textsb{83.8} & \textsb{52.0} & 96.4 & \textsb{76.8}\\
    \midrule
    \textsb{Hybrid BYOL-S/CvT:}\\
    \textsb{$\alpha$:$\beta$ ratio}\\
    1:4 & -$^*$ & 72.5          & 42.7          & 88.7          & 68.0\\
    1:2 & -$^*$ & 78.4          & 46.1          & 90.5          & 71.7\\
    2:3 & -$^*$ & 79.6          & 47.8          & 96.4          & 74.6\\
    1:1 & -$^*$ & \textsb{83.8} & \textsb{50.2} & 96.4          & \textsb{76.8}\\
    3:2 & -$^*$ & 82.3          & 47.9          & \textsb{97.6} & 75.9\\
    2:1 & -$^*$ & 83.1          & 48.3          & 95.8          & 75.7\\
    4:1 & -$^*$ & 82.4          & 48.7          & 94.0          & 75.0\\
    \bottomrule
  \end{tabular}
  \label{tab:env}\vspace{-0.25cm}
\end{table}
\setlength{\tabcolsep}{2.5pt}
\begin{table}[!htb]
    \small
    \centering
    \caption{Top-1 accuracy of all proposed models on music-related tasks. S/M: Speech vs. Music task.}
    \begin{NiceTabular}{l@{\hspace{10pt}}c@{\hspace{10pt}}*{2}{c}@{\hspace{10pt}}*{2}{c}@{\hspace{10pt}}*{2}{c}@{\hspace{10pt}}*{2}{c}@{\hspace{10pt}}c}
    \toprule
    \rule{0pt}{65pt} & \Block{1-1}{\rotate Beijing Opera} & \Block{1-2}{\rotate GTZAN} && \Block{1-2}{\rotate Mridangam} &&  \Block{1-2}{\rotate NSynth (5h)} && \Block{1-2}{\rotate NSynth (50h)} && \rotatebox{90}{Average} \\
    & & Genre & S/M & Stroke & Tonic & Pitch & Chroma & Pitch & Chroma \\
    \midrule
    \textsb{HEAR baselines:} \\
    CREPE & 93.2 & 64.5 & 86.7 & 88.7 & 82.3 & \textsb{87.2} & \textsb{93.4} & \textsb{89.5} & \textsb{95.2} & \textsb{86.7} \\
    wav2vec2 & 89.4 & 78.0 & 92.3 & 94.7 & 82.8 & 40.0 & 44.6 & 66.7 & 71.9 & 73.4 \\
    \midrule
    \textsb{Pre-training dataset:}\\
    BYOL-A   & 91.9 & 83.5 & \textsb{96.9}  & 97.0   & 90.0  & 29.0  & 36.0 & 64.2 & 67.5 & 72.3  \\
    BYOL-S & 91.1  & 83.8 & 92.3  & 97.3  & \textsb{92.9} & 42.0 & 44.0 & 70.0 & 73.4 & 76.3 \\
    BYOL-S\texttt{++} & \textsb{95.3}  & \textsb{83.9} & 93.8 & \textsb{97.4} & 91.7 & 40.0 & 41.8 & 69.6 & 73.2 & 76.3 \\
    \midrule
    \textsb{BYOL-S window size (s):}\\
    0.5   & \textsb{94.9} & 83.5 & 96.9 & \textsb{97.3}           &  \textsb{93.0} & 38.2           &  40.6           &  71.2          &  74.5 & 76.7           \\
    
    0.95  & 93.2          & 82.5 & 96.2 & \textsb{97.3} & 92.5 & 38.2            & 39.0            & 69.0           & 72.1 & 75.6                 \\
    1.425 & 94.5    & \textsb{83.6} &   \textsb{98.5}      & 97.0          & 92.5    & 35.2            & 37.6           & 67.7          & 71.6 & 75.4                  \\
    2     & 94.1       & 83.4 &       \textsb{98.5} &  97.1           & 91.9   & 35.8            & 37.4           & 66.2          & 69.2 & 74.8  \\
    \midrule
    \textsb{BYOL-S encoder:}\\
    Default  & 91.1 & 83.8 & 92.3 & 97.3 & 92.9 & 42.0 & 44.0 & 70.0 & 73.4 & 76.3 \\
    Resnetish-34 & 92.4 & 77.4 & 96.9 & 96.1  & 88.7 & 26.0 & 22.4 & 44.9 & 47.4 & 65.8 \\
    CLSTM & \textsb{97.0} & 75.9 & 96.2 & 96.7 & 89.7 & 28.6 & 32.0 & 57.6 & 65.0 & 71.0 \\
    CvT & 96.2 & \textsb{84.2} & \textsb{97.7} & \textsb{97.5} & \textsb{94.0} & 53.0 & 55.2 & 76.1 & 85.0 & 82.1 \\
    \midrule
    \textsb{Hybrid models:} \\
    openSMILE (OS) only & 89.4 & 78.2 & \textsb{97.6} & 95.6 & 88.0 & 47.0 & 51.0 & 73.9 & 78.3 & 77.7\\
    BYOL-S/CvT + OS & 92.8 & 83.4 & 96.9 & 96.8 & 93.1 & 53.4 & 56.6 & 80.9 & 85.6 & 82.2\\
    Hybrid BYOL-S/Default   & 92.8 & \textsb{85.9} & 96.9 & 96.1 & 88.7 & 35.8 & 38.0  & 71.1 & 75.0 & 75.6  \\
    Hybrid BYOL-S/CvT       & \textsb{94.5} & 85.8 & 96.2 & \textsb{97.7} & \textsb{94.9} & 56.8 & 59.2 & 79.5 & 83.0 & 83.1  \\
    \midrule
    \textsb{Hybrid BYOL-S/CvT:}\\
    \textsb{$\alpha$:$\beta$ ratio}\\
    1:4 & 94.9 & 82.2 & 95.3 & 97.2 & 96.0 & 66.4 & 68.6 & 80.8 & 84.6 & 85.1\\
    1:2 & \textsb{96.2} & 84.8 & 94.6 & 97.6 & \textsb{96.7} & 65.4 & 69.6 & 81.6 & 85.8 & 85.8\\
    2:3 & 93.2 & 86.1 & 93.8 & 97.5 & 96.6 & 67.0 & 69.8 & 81.1 & 85.4 & 85.6\\
    1:1 & 94.5 & 85.8 & \textsb{96.2} & 97.7 & 94.9 & 56.8 & 59.2 & 79.5 & 83.0 & 83.1\\
    3:2 & 95.8 & 85.6 & 94.6 & 97.7 & \textsb{96.7} & 64.0 & 67.0 & 81.9 & 85.9 & 85.5\\
    2:1 & 94.9 & 86.4 & 94.6 & \textsb{97.8} & 96.6 & 62.8 & 66.4 & 80.9 & 85.0 & 85.0\\
    4:1 & 94.9 & \textsb{86.8} & 92.3 & \textsb{97.8} & 96.4 & 63.2 & 65.6 & 79.9 & 84.3 & 84.6\\
    \bottomrule
    \end{NiceTabular}
    \label{tab:music}
\end{table}
\setlength{\tabcolsep}{6pt}

\clearpage
\subsection{Timestamp Embeddings}
In this section, we evaluate our models on tasks that depend on generating embeddings from segments, instead of the entire audio recording, to detect sound events—onset and offset—or music transcription. We tested this format on the two timestamp-based tasks from the HEAR challenge: DCASE and MAESTRO (Table \ref{tab:timestamp_data}). For the best model (hybrid BYOL-S/CvT), two hyperparameters were tuned: the window size and the hop size. The evaluation results are presented in Tables~\ref{tab:timestamp_main} and \ref{tab:timestamp_window}.

\setlength{\tabcolsep}{3pt}
\begin{table*}[h!]
    \small
    \centering
    \vspace{-0.5cm}
    \caption{Results on timestamp-based tasks. All models used input windows of 1 s with a 50 ms hop size. $\downarrow$ indicates lower is better.
    Due to exhausting GPU memory problems for openSMILE features with MAESTRO (shown with superscript $^*$), no reported average for these features.}
    \begin{tabular}{l*5c}
    \toprule
    & \multicolumn{2}{c}{DCASE} & \multicolumn{2}{c}{MAESTRO}  & Average \\
    & Onset FMS & Error rate $\downarrow$ & Onset FMS & Onset w/ Offset FMS & Onset FMS\\
    \midrule
    \textsb{HEAR baselines:} \\
    CREPE & 0.552 & 0.420 & \textsb{0.3910} & \textsb{0.15} & 0.472 \\
    wav2vec2 & 0.670 & 0.320 & 0.0328 & 0.009 & 0.351 \\
    \midrule
    \textsb{Models:} \\
    BYOL-A &  0.499 & 0.503 & 0.0028 & 0.00029 & 0.251\\
    BYOL-S &  0.650 & 0.356 & 0.0043 & 0.00048 & 0.327 \\
    BYOL-S\texttt{++} & 0.512 & 0.583 & 0.0457 & 0.01055 & 0.279 \\
    Hybrid BYOL-S & 0.526 & 0.504 & 0.0067 & 0.00090 & 0.266 \\
    BYOL-S/CvT & \textsb{0.891} & \textsb{0.152} & 0.0817 & 0.01488 & \textsb{0.486} \\
    openSMILE (OS) only & 0.857 & 0.194 & -$^*$ & -$^*$ & n/a \\
    BYOL-S/CvT + OS & 0.865 & 0.18 & -$^*$ & -$^*$ & n/a \\
    Hybrid BYOL-S/CvT & 0.889 & 0.153 & 0.0746 & 0.01242 & 0.482 \\
     \bottomrule
    \end{tabular}
    
    \label{tab:timestamp_main}
    \end{table*}
\setlength{\tabcolsep}{3pt}
\begin{table*}[h!]
    \small
    \centering
    \vspace{-0.5cm}
    \caption{Results on timestamp-based tasks using the hybrid version of BYOL-S/CvT. $\downarrow$ indicates lower is better.}
    \begin{tabular}{l*5c}
    \toprule
    & \multicolumn{2}{c}{DCASE} & \multicolumn{2}{c}{MAESTRO} & Average \\
    & Onset FMS & Error rate $\downarrow$ & Onset FMS & Onset w/ offset FMS & Onset FMS\\
    \midrule
    \textsb{HEAR baselines:} \\
    CREPE & 0.552 & 0.420 & \textsb{0.3910} & \textsb{0.15} & 0.472 \\
    wav2vec2 & 0.670 & 0.320 & 0.0328 & 0.009 & 0.351 \\
    \midrule
    \textsb{Hybrid BYOL-S/CvT:} \\
    \textsb{Window/Hop size} \\
    1s/50ms & \textsb{0.889} & 0.153 & 0.0746 & 0.01242 & 0.482 \\
    0.5s/50ms & 0.880 & \textsb{0.144} & 0.1739 & 0.04406 & \textsb{0.527} \\
    \bottomrule
    \end{tabular}
    \label{tab:timestamp_window}
\end{table*}
\setlength{\tabcolsep}{6pt}

\section{Discussion}

\subsection{Contribution of pre-training datasets: BYOL-A, BYOL-S and BYOL-S\texttt{++}}

Table~\ref{tab:speech} shows that BYOL-S\texttt{++} outperformed BYOL-A and BYOL-S in all speech-related tasks. This result is consistent with the fact that BYOL-S\texttt{++} was trained on a larger and more diverse speech corpus, with both spontaneous and anechoic scripted speech. In addition, BYOL-S surpassed the other models in all environmental sound tasks as well as in most music-related tasks (Tables~\ref{tab:env}, \ref{tab:music}). The latter observation comes as less intuitive: one could expect BYOL-A, a general-purpose audio representation, to perform best on these tasks, whereas BYOL-S is speech-specific. However, it is worth mentioning that the speech-subset of AudioSet included speech utterances with various background setups (e.g. music, environmental sounds, vocalizations, etc.). Accordingly, we hypothesize that the high performance of BYOL-S on non-speech tasks might be due to the model’s exposure to both speech and non-speech during training.


On the other hand, BYOL-A performed significantly better than the other models in differentiating music from speech (GTZAN; Table \ref{tab:music}), which could be due to being trained on all AudioSet ontologies and thus acquiring better discrimination of signal types.

Unsurprisingly, the HEAR baseline CREPE~\citep{kim2018crepe} consistently outperformed the proposed approaches in the NSynth pitch discrimination task. Indeed, CREPE is a specialized pitch representation, rather than a general-purpose audio model. The performance gap might be explained by different pitch properties in speech and instrumental music. Since the proposed model is predominantly a speech feature extractor, it does not fully capture pitch representation in musical tasks. However, the performance gap seems to decrease for larger downstream dataset sizes (i.e., NSynth (50h); Table \ref{tab:music}).

To gain further insight into model performance, we used the centered kernel alignment (CKA) method \citep{cortes2012, kornblith2019} to assess the similarity between layers of the BYOL-A, BYOL-S, and BYOL-S\texttt{++} representations (Figure \ref{fig:cka_plots}). Following \citet{raghu2021vision} and \citet{subramanian2021torch_cka}\footnote{\url{https://github.com/AntixK/PyTorch-Model-Compare}}, we first fed an unseen dataset (CREMA-D) to two different models (e.g., BYOL-A and BYOL-S\texttt{++}; Figure \ref{fig:cka_plots}, left) to generate the activation maps for each layer. Using the CKA algorithm, we were then able to compute pairwise similarity scores between each layer using the Gram matrix of their activation maps, resulting in a “similarity matrix" between the two models. 

\begin{figure}
    \floatconts
    {fig:cka_plots}
    {\caption{CKA-based model similarity analysis.}}
    {
        \includegraphics{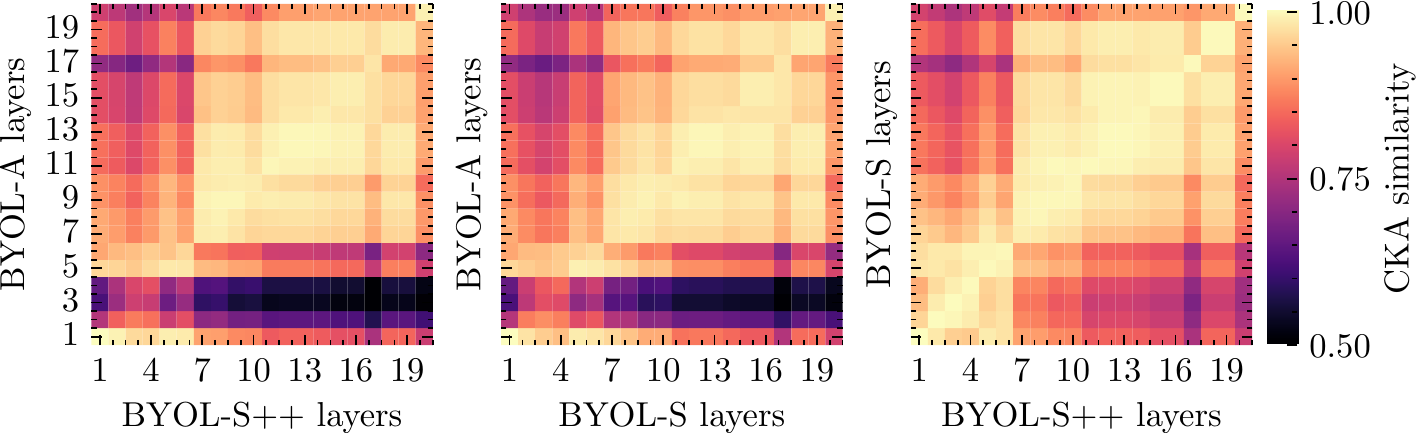}
    }
\end{figure}

Accordingly, Figure \ref{fig:cka_plots} shows that only the last convolution layers of BYOL-A shared a high CKA similarity with those of BYOL-S and BYOL-S\texttt{++}, constituting reasonable evidence that BYOL-A learned different features from the speech-specific frameworks. Conversely, high similarity was observed across all layers between BYOL-S and BYOL-S\texttt{++}. This observation is consistent with Tables \ref{tab:speech}-\ref{tab:music}, where the BYOL-S and BYOL-S\texttt{++} achieved similar results on the HEAR benchmark. Thus, the lower similarity between the first layers of BYOL-A and BYOL-S-derived models could mean that these first layers have a detrimental influence on model performance.
More generally, these results confirm that pre-training dataset selection is, unsurprisingly, critical to produce robust audio representations, especially as all models presented here shared the speech portion of AudioSet for pre-training.

\subsection{Contribution of the pre-training window size}
Table \ref{tab:speech} depicts a trend towards higher performance in speech-related scene-based tasks for larger window sizes during pre-training. This influence of the training window size is not unreasonable since a typical utterance can span across several seconds. Conversely, for environmental sounds and music (comprising transient musical notes), Tables \ref{tab:env} and \ref{tab:music} reflect a reverse tendency, i.e., the smaller the window size the better in most tasks. This might be due to the fact that brief environmental sounds (r.g., gunfire sounds from the Gunshot triangulation dataset or glass breaking and dog barking in the ESC dataset) do not require a large context window to be identified, making the models that were pre-trained on a small context window a better fit. As window size optimization appears to be highly dependent on the dataset, developing training protocols to effectively capture context across multiple timescales should constitute for a crucial step to produce universal audio representations.

\subsection{Comparison of encoder architectures: Resnetish-34, CLSTM, CvT}
Among the alternative encoders to the default CNN-based encoder network in BYOL-S (Section \ref{sec:encoder}), only BYOL-S/CvT outperformed the “vanilla" BYOL-S. For instance, BYOL-S/CvT was the best representation for music-related tasks (Table  \ref{tab:music}), and achieved similar performance to BYOL-S in speech-related tasks (Table \ref{tab:speech}). On the other hand, the default BYOL-S outperformed all other encoders in environmental sound-related tasks (Table \ref{tab:env}). To explain such discrepancies, one could hypothesize that using the other encoders, CLSTM and ResNetish34, caused overfitting problems during pre-training due to their higher model capacity. This hypothesis is especially motivated by the fact that both frameworks had a remarkably lower pre-training error compared to BYOL-S and BYOL-S/CvT. Thus, one could argue that simple encoder models might be preferable for generating robust audio representations when using the BYOL-A paradigm. 
That being said, larger transformer-based models and trained on large-scale datasets tend to yield superior audio representations~\citep{shor2021universal}. However, due to their size, the application of such models is costly and necessitates a longer inference time. Here, we focused on comparatively smaller models ($<$ 30M parameters), which can be trained and applied using manageable computing resources. 

\subsection{Hybrid versions of BYOL-S and BYOL-S/CvT}
In this study, we proposed \textit{hybrid} models of BYOL-S as an attempt to take advantage of DSP-based features and eventually increase feature interpretability for data-driven features. In fact, the hybrid version of BYOL-S/CvT outperformed the other proposed models on most tasks of the HEAR benchmark (Tables \ref{tab:speech}-\ref{tab:music}). In particular, the benefit of this hybrid method was especially apparent when utilizing CvT as the encoder. Hence, adding DSP-based fixed features could be viewed as a good auxiliary task to support model pre-training. Consequently, at the end of the pre-training, the model should be able to strike the optimal trade-off between the pure-DSP and a fully data-driven, learned representation. The addition of fixed, non-trainable features could also improve training stability (i.e., preventing model collapse), which is known to be one of the issues associated with BYOL-style methods (usually mitigated by updating the target weights as a moving average of those of the online network) \citep{grill2020}. While the supervision module (Figure~\ref{fig:hybrid_byols_plot}) only consisted of fixed DSP-based features from openSMILE in this work, using other learnable, deep learning-based, feature extractors in the supervision module could constitute promising future work, where the two trainable systems could be even trained or fine-tuned in an iterative fashion. 
To explore the impact of the supervised and self-supervised components on hybrid model performance, we varied the weights $\alpha$ and $\beta$ given to their respective loss functions. Overall, the results of this tuning procedure seem to be dataset-dependent. In speech and environmental sound tasks (Tables \ref{tab:speech} and \ref{tab:env}), performance was maximal when $\alpha = \beta =$ 1 and gradually decreased as the ratios become more unbalanced. In music-related tasks (Table \ref{tab:music}), using $\alpha = \beta =$ 1 yielded notably worse results than for other ratios.

To complement our experiments, we assessed the validity of the hybrid training protocol by comparing the hybrid version of BYOL-S/CvT with embeddings produced using only openSMILE features or the concatenation of the latter with BYOL-S/CvT embeddings. As shown in Tables \ref{tab:speech}-\ref{tab:music}, we found that the hybrid model remained the best representation, suggesting that the lower performance for the concatenated features might be an example of the curse of dimensionality. Increasing the dimensions of the feature vector requires more complex decision boundaries for classifiers. Also, adding more features increase the likelihood of having correlated features which might impact some learning algorithms. Thus, we argue that the information learned during the hybrid training paradigm is more robust than a simple concatenation between features having different distributions while maintaining the same feature size.



\subsection{Timestamp Embeddings}
Table \ref{tab:timestamp_main} showed that the CvT encoder and the hybrid models constituted on average our best performing models for timestamp-based tasks. It is noted that the performance of openSMILE features only, on the DCASE task, was improved by concatenating the latter features with BYOL-S/CvT, however, Hybrid BYOL-S/CvT still outperforms the simple concatenation between both features. While yielding marginal changes for the DCASE task (Table \ref{tab:timestamp_window}), decreasing the window size for the hybrid BYOL-S/CvT considerably improved the detection of note onsets in MAESTRO, in addition to note onset with offset frames. This could be because music notes tend to be short transient events, making small window lengths a better choice for music transcription tasks. That being said, our current implementation for timestamp embeddings remained a relatively simple extension of that for scene embeddings. Hence, it is likely that a specialized encoder, or an entire different pre-training protocol, could yield substantially better results. That is why HEAR's implementation of CREPE, specifically designed as pitch representation, as well as other CREPE-derived models~\citep{turian2022}, consistently topped the leaderboard on musical tasks.

\section{Conclusion}
In this paper, we present our submission for the 2021 NeurIPS HEAR challenge, a benchmark to evaluate audio representations on 16 downstream tasks. Our submission was based on BYOL-S, a re-trained version of BYOL-A on a speech subset of AudioSet. Although the model was originally designed as a speech-specific representation, we showed that it can also produce suitable performance in other tasks involving environmental sounds and music. Additionally, we carried out several experiments to delve deeper into the model components that contribute to generating audio representations. The experiments included trying different pre-training datasets, encoder architectures, pre-training audio windows, a hybrid pre-training protocol, and finally, the selection of hyperparameters to optimize timestamp embeddings. While the choice of pre-training audio size was generally subject-dependent (e.g., shorter and longer windows for environmental sounds and speech, respectively), opting for CvT-based encoding and a hybrid training protocol using openSMILE features yielded more robust results. Consequently, we observed that the hybrid version of BYOL-S/CvT, i.e., with CvT encoding and hybrid pre-training, outperformed our original HEAR challenge submission (BYOL-S), constituting our best audio representation with respect to the HEAR benchmark. Additionally, this approach was evaluated on detecting cognitive and physical load from speech samples showing competitive results against state-of-the-art speech representations \citep{elbanna22_interspeech}. Thus, combining original self-supervised pre-training paradigm with DSP-based handcrafted features used in the loss function of the hybrid model helped produce a more robust audio representation. This finding further validates the benefit of using ensemble embeddings obtained from several models for general-purpose audio representation~\citep{turian2022}. In particular, involving a considerably simple set of fixed features during training can substantially improve DNN-based audio representation learning. For future work, it would be worth experimenting with different augmentation methods in time and frequency domains and exploring more encoder architectures to improve the robustness of the learnt representations.

\begingroup
\setstretch{0.98}
\bibliography{references.bib}
\endgroup

\let\clearpage\relax

\end{document}